\documentstyle[11pt,newpasp,twoside,epsf]{article}
\def\edcomment#1{\iffalse\marginpar{\raggedright\sl#1\/}\else\relax\fi}
\reversemarginpar
\begin{document}
\title{Rapid Variability as a Diagnostic of Accretion and Nuclear
Burning in Symbiotic Stars and Supersoft X-ray Sources} 
 \author{J. L. Sokoloski} 
\affil{Smithsonian Astrophysical Observatory, MS 15, 60 Garden St.,
Cambridge, MA 02138}

\begin{abstract}
Accretion disks and nuclear shell burning are present in some
symbiotic stars (SS) and probably all supersoft X-ray binaries
(SSXBs).  Both the disk and burning shell may be involved in the
production of dramatic outbursts and, in some cases, collimated jets.
A strong magnetic field may also affect the accretion flow and
activity in some systems.  Rapid-variability studies can probe the
interesting region close to the accreting white dwarf (WD) in both SS
and SSXBs.  I describe fast photometric observations of several
individual systems in detail, and review the results of a photometric
variability survey of 35 SS.  These timing studies reveal the first
clearly magnetic SS (Z And), and suggest that an accretion disk is
involved in jet production in CH Cyg as well as in the outbursts of
both CH Cyg and Z And.  They also support the notion that the
fundamental power source in most SS is nuclear burning on the surface
of a WD, and raise questions about the structure of disks in the
SSXBs.  Finally, spectroscopic observations of RS Oph reveal
minute-time-scale line-strength variations, probably due to a hot
boundary layer.  Taken together, the rapid timing observations explore
the connections between jet-producing WDs and X-ray binaries, as well
as SS, SSXBs, and CVs.
\end{abstract}

\section{Introduction}

Accretion onto a white dwarf (WD) and nuclear shell burning probably
power the observed activity in supersoft X-ray binaries (SSXBs) and
many symbiotic stars (SS).  Besides the hallmark high-excitation-state
emission lines (for SS) and soft X-ray spectra (for SSXBs), this
activity can include outbursts and collimated jets.  But in SS,
emission and absorption by the nebula can hide the spectroscopic
signatures of a disk and absorb the soft X-rays from the
nuclear-burning shell.  In SSXBs, the disk is optically visible, but
the optical emission is dominated by reflected nuclear-burning light,
and the disk is significantly heated (e.g., Popham \& Di Stefano
1996).  One way to examine the region close to the accreting WD
despite these complications is to look for rapid variations.

In the context of this work, `rapid' or `fast' variations are those
which could be associated with WD-disk phenomena.  Stochastic variations
from a disk, termed ``flickering'', generally occur on either a dynamical time
\begin{eqnarray*}
t_{dyn} & \sim & \frac{1}{\Omega_k} \sim 4\,{\rm s}\; \left(
\frac{r}{10^{9} {\rm cm}} \right)^{3/2} \left( \frac{M_{WD}}{0.6\,M_{\odot}}
\right)^{-1/2}
\end{eqnarray*}
(where $\Omega_k = \Omega_k(r)$ is the Keplerian angular velocity, $r$
is the radial 
position in the disk, and $M_{WD}$ is the mass of the WD), or a
viscous time
\begin{displaymath}
t_{visc}  \sim  \frac{1}{\alpha} \left( \frac{r}{H} \right)^2
\frac{1}{\Omega_k} 
 \sim  8\, {\rm hr}\; \left( \frac{\alpha}{0.1} \right)^{-4/5}
\left( \frac{\dot{M}}{10^{-8} M_{\odot} \ {\rm yr^{-1}}} \right)
^{-3/10} 
\left( \frac{r}{10^{9} {\rm cm}} \right)^{5/4} \\
\end{displaymath}
(where $H$ is the disk height, $\alpha$ is the viscosity parameter,
and $\dot{M}$ is the accretion rate onto the WD; Frank, King, and
Raine 1992).  Disk flickering therefore depends on the size and
properties of the disk, and the location of the emitting region within
the disk.  In WD disks, fluctuations occur with time scales of roughly
a day or less, and observations verify that the fastest stochastic
variations come from the inner disk (e.g., Bruch 2000).

Brightness oscillations due to magnetic accretion should also have
time scales ranging from minutes to hours.  When a WD is in spin
equilibrium (neither being spun up nor spun down by torques from the
disk material), it rotates with the Keplerian frequency at the radius
where the ram pressure from in-falling material balances the pressure
in the magnetic field.  For fields typically measured in intermediate
polars (IPs; $B_S \sim 10^5 - 10^6$ G) and accretion rates that are
reasonable for SS or SSXBs ($\dot{M} \sim 10^{-9} - 10^{-7}
{M}_{\odot} \ {\rm yr^{-1}}$), the equilibrium spin periods are tens
of minutes.

Changes in the luminosity from a burning shell
occur on the nuclear time, $t_{nuc}$, which is roughly the time to
accrete the envelope scaled by the fraction of the envelope that must
be burned to heat the entire envelope to temperature $T$:
\begin{eqnarray*}
t_{nuc} & \sim & \left( \frac{C_P T}{E_{nuc}} \right) \left( \frac{\Delta
M}{\dot{M}} \right) \\ 
 & \sim &  3\, {\rm yr}\; \left( \frac{T}{3\times 10^7 {\rm K}} \right)
\left( \frac{\Delta M}{6 \times 10^{-5} M_{\odot}} \right) 
\left( \frac{\dot{M}}{4\times 10^{-8} M_{\odot} \ {\rm yr^{-1}}}
\right)^{-1}, \\ 
\end{eqnarray*}
where $C_P$ is the specific heat at constant pressure, $E_{nuc}$ is the energy released
per gram from nuclear burning of H, and $\Delta M$ is
the mass of the envelope, which is dependent on $M_{WD}$ and
$\dot{M}$ (Fujimoto 1982).  
Therefore, fast variations cannot be due to fundamental changes in the
nuclear burning emission (changes due to obscuration or reflection of
this emission, however, could happen quickly).

Even though the nuclear burning luminosity cannot change quickly, it
is still an important consideration in rapid-variability studies.
Since $E_{nuc}/E_{grav} \approx 40$ (where $E_{grav}$
is the energy released per gram from accretion onto the WD),
quasi-steady nuclear shell burning dominates the energetics of the hot
component when present.  Nuclear burning can therefore hide or
diminish rapid variations that are directly associated with accretion.
The maximum and minimum accretion rates to produce quasi-steady shell
burning are
\begin{eqnarray*}
\dot{M}_{steady,max} &  = & 2.8\times 10^{-7} + \; 5.9 \times 10^{-7}
\left( \frac{M_{WD}}{M_{\odot}} -1.0 \right) M_{\odot} \ {\rm yr^{-1}} \\
\dot{M}_{steady,min} & = & 1.32 \times 10^{-7} M_{WD}^{3.57}
M_{\odot} \ {\rm yr^{-1}}. 
\end{eqnarray*}
(Paczy{\'n}ski \& Rudak 1980; Iben 1982).  If $\dot{M} >
\dot{M}_{steady,max}$, the fuel cannot be burned as fast as it is
accreted, and the envelope could expand or be ejected.  If $\dot{M} <
\dot{M}_{steady,min}$, material burns unstably in nova explosions,
although a period of residual burning may follow.  If the
hot-component luminosity ($L_{hot}$) is greater than approximately
$100\,L_{\odot}$, significant nuclear burning must be present, since
such high luminosities cannot be produced by accretion with $\dot{M} <
\dot{M}_{steady,min}$.

One final time scale of interest is the recombination time in a
SS nebula,
\begin{eqnarray*}
 t_{rec} & \sim & \frac{1}{n_s \alpha_B} \sim 1 {\rm hr}\;
\left( \frac{n_s}{10^{9}\, {\rm cm}^{-3}} \right)^{-1} \left(
\frac{\alpha_B}{2.59 \times 10^{-13}\, {\rm cm}^3/{\rm s}}
\right)^{-1}\;\;\; {\rm at}\; 
10^4\,{\rm K} 
\end{eqnarray*}
(where $n_s$ is the density at the outer edge of the ionized region
and $\alpha_B$ is the case-B recombination coefficient;
Fern{\'a}ndez-Castro et al. 1995).  Any high-energy (far-UV or soft
X-ray) variations faster than $t_{rec}$ will be smeared out 
when the nebula reprocesses them into the optical.

In \S2, I discuss the first known case of a WD with both magnetically
channeled accretion and surface nuclear burning (Z And).  I describe
photometric evidence for changes in an accretion disk associated with
the production of a jet in a WD accretor (CH Cyg) in \S3, and discuss
how high-time-resolution studies can reveal the relative importance of
nuclear burning vs. viscous dissipation in \S4.  Disk flickering from
the supersoft source MR Vel is described in \S5, 
and minute-time-scale spectral line variability that can act as a
diagnostic for the physical conditions in the line-emitting regions
(RS Oph) is shown in \S6.  I discuss the overall results from these
variability studies in \S7.

\section{Magnetic Accretion in Z Andromedae} \label{sec:zand} 

In IPs, the WD magnetic field is strong enough to truncate the inner
accretion disk, channel material onto the WD polar caps, and produce
an X-ray and/or optical oscillation at the WD spin period (see e.g.,
Warner 1995).  If strong WD fields are fossil fields of magnetic Ap
and Bp stars (see Wickramasinghe \& Ferrario 2000), then the fraction
of magnetic WDs in interacting binaries should be higher than the
fraction of single WDs with strong fields\footnote{Because the
accreting WD in a CV-like interacting binary was originally the more
massive of the pair, the progenitor population of these WDs contains
relatively more Ap and Bp stars.}.  A higher fraction of WDs in CVs do
appear to be magnetic (25\% vs. 5\% in the field; Wickramasinghe \&
Ferrario 2000).  But selection effects could bias this comparison,
since CVs are often discovered because of their X-ray pulsations.  It
is therefore useful to examine different classes of WD accretors.
Until recently, however, no IP-like oscillations had been found in
either a SS or SSXB.

\begin{figure}[t]
\plottwo{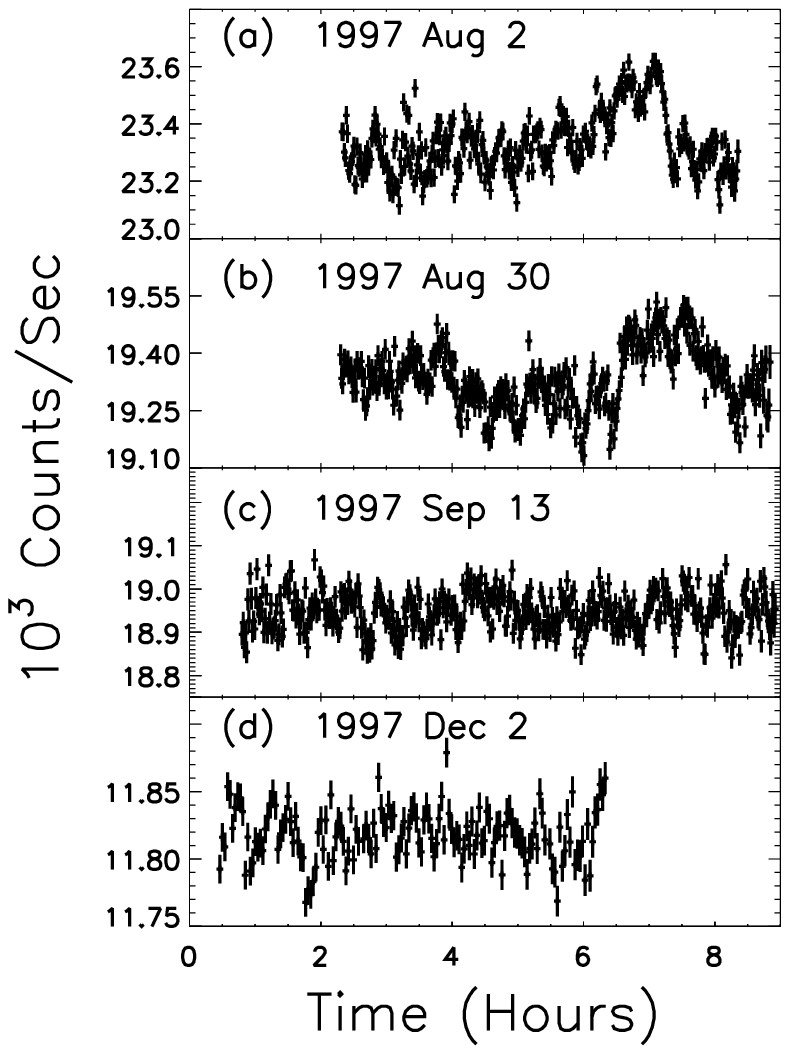}{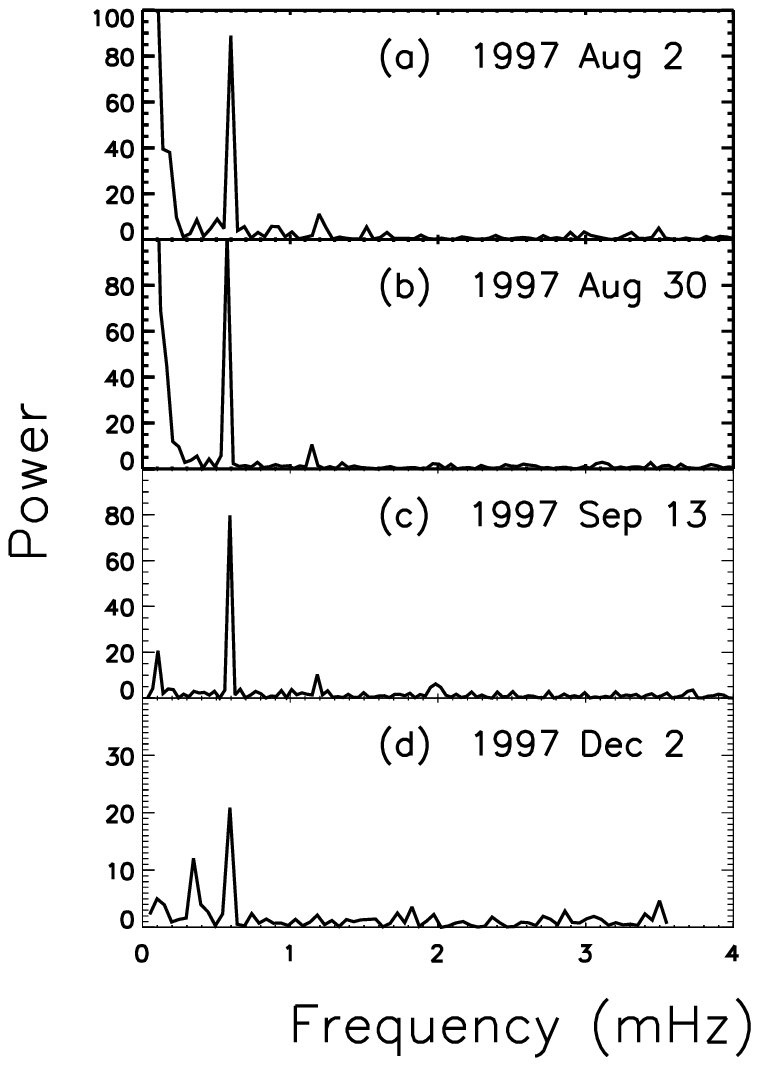} 
\caption{Example Z And $B$-band light curves and corresponding power
spectra (from Sokoloski \& Bildsten 1999).  
The power is plotted in units of mean high frequency power. The
feature at 0.6 mHz corresponds to an oscillation period of 28 m.}
\end{figure}

In 1997 and 1998, Sokoloski \& Bildsten (1999) observed the
prototypical symbiotic Z And on 8 occasions, with the 1-m telescope at
Lick Observatory, and found a 28-minute oscillation each time.  Figure
1 shows example $B$-band light curves and the corresponding power
spectra.  Taking reasonable values for the WD mass, radius, and
accretion rate, a spin period of 28 m implies a WD surface magnetic
field strength of at least $B_S = 3
\times 10^4$ G, and if the WD is in spin equilibrium, 
$B_S = 6 \times 10^6$ G (Sokoloski \& Bildsten 1999).  This field is
comparable to fields found in magnetic CVs. The amplitude of the
optical oscillation ($<2$ mmag, or $< 0.2$\% in quiescence), however,
is roughly 10 to 100 times smaller than the optical spin modulations
seen in magnetic CVs.  If such low oscillation amplitudes are typical
of magnetic SS, then they were probably not discovered earlier because
past observations were not sufficiently sensitive.

\begin{figure}
\plottwo{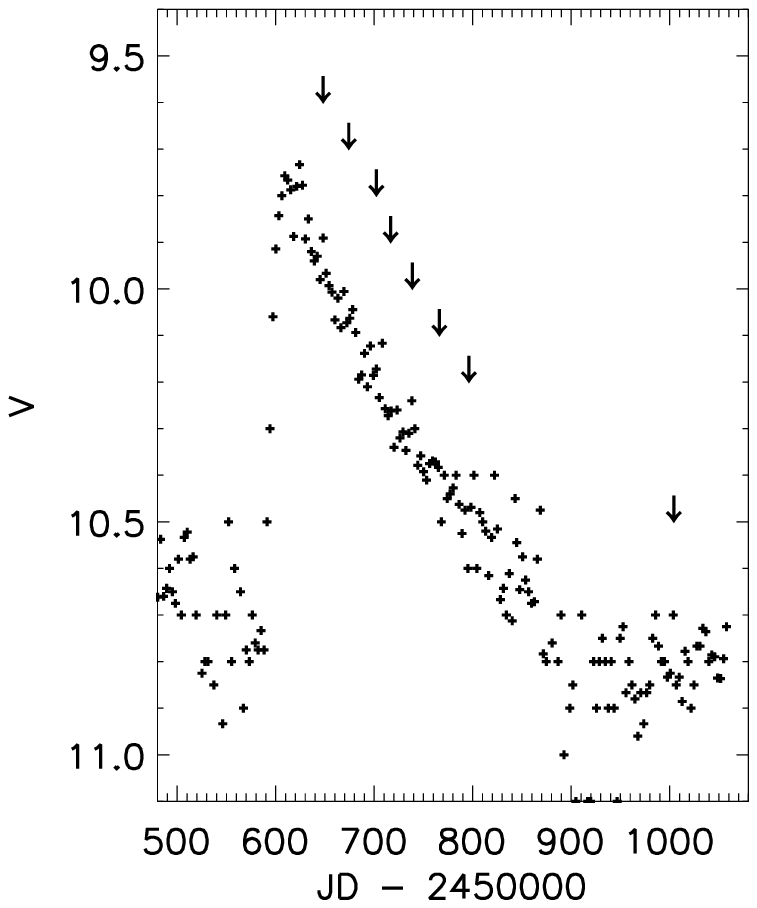}{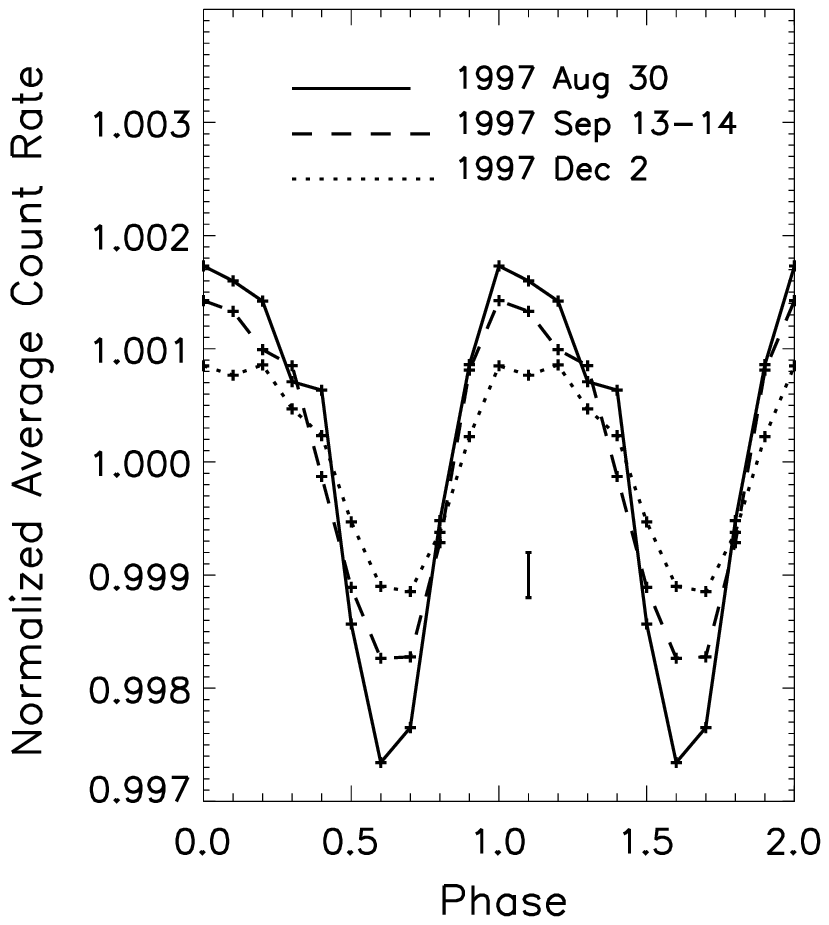} 
\caption{Z And long-term $V$-band light curve with the times of the 8
fast photometric observations marked with arrows (left), and the pulse
profile evolution (right; from Sokoloski \& Bildsten 1999).
The size of a typical 1$\sigma$ error is shown in the center of the
pulse profile plot.}
\end{figure}

M\"{u}rset et al. (1991) estimate a WD luminosity for Z And of
$L_{hot} \approx 600 - 1600 L_{\odot}$, so Z And is probably
experiencing some nuclear shell burning.  Two important points follow:
1) magnetic accretion can produce an optical modulation at the WD spin
period even in a WD with quasi-steady nuclear burning (and Z And is
the first example), and 2) the presence of nuclear shell burning and a
nebula that can reprocess high-energy emission into the optical may
reduce the amplitude of the optical modulation.

Osborne et al. (2001) found what may be the first X-ray oscillation
due to magnetic accretion onto a WD with surface nuclear burning - in
the supersoft source M31PSS.  The X-ray
modulation amplitude was roughly 50\%.  King, Osborne,
\& Schenker (2002) 
argue that the X-ray flux varies at the WD spin period because the
nuclear burning rate is higher near the poles, where the burning shell
is preferentially supplied with fresh fuel.
In contrast, the X-ray flux of Z And was constant to within the errors
in an XMM observation on 2001 28 January (A. Kong, private
communication).
The amplitude of any sinusoidal spin modulation
was less than roughly 10\%.
The small X-ray oscillation amplitude compared to M31PSS could mean
that the high WD luminosity in Z And is due to residual burning from a
previous nova rather than accretion with $\dot{M} >
\dot{M}_{steady,min}$.  
Alternatively, the X-ray oscillation amplitude could be reduced by the
presence of an additional source of X-ray flux, such as emission from
shock-heated colliding winds.

The behavior of the 28-m optical oscillation amplitude during two  
recent outbursts of Z And has implications for the nature of these
events.  In 1997, the  
oscillation amplitude was highest near the optical peak of the
outburst, and decreased as the outburst faded (see Figure 2).  The hot
spots near the WD surface were therefore not hidden by an expanded
photosphere.  Instead, the oscillation-amplitude evolution is
consistent with an increase in $\dot{M}$
during outburst, such as from a dwarf-nova-like disk instability.
Near the optical peak of the larger outburst in 2000-2002, on the
other hand, the optical oscillation was not detected, and FUV
observations indicate that a shell of material was ejected (Sokoloski
et al. 2002). The 2000 outburst may have also been triggered by an
disk instability, but with a more dramatic response by the burning
layer.

\vspace{-0.2cm}
\section{Disk-Jet Connection in CH Cygni} \label{sec:chcyg}

In CH Cyg, both the optical flickering amplitude and power-spectrum
shape can change.  Figure 3 shows some example high-speed light
curves, and Figure 4 (top) shows the long-term light curve with the
times of the fast photometric observations marked.  In 1996 and 1997,
CH Cyg dropped to a very low optical state, and low amplitude, rolling
(hour-time-scale) variations were present, with little to no
minute-time-scale variability (see also Rodgers et al. 1997).  This
type of smooth light curve is unusual for CH Cyg, which more commonly
shows CV-like fluctuations on time scales from seconds to hours.  In
1997 August, when the optical flux began to rise again, the
minute-time-scale variations returned, although the overall flickering
amplitude was still low ($\la 0.1$ mag).  In 1998 August, when CH Cyg
returned to an optical high state, the peak-to-peak flickering
amplitude increased to $\la 0.5$ mag.  In 1999 July, the optical flux
was constant to within the detection limit of a few mmag, presumably
due to an eclipse of the accreting WD by a companion orbiting with a
period of roughly 14 yr (e.g., Crocker et al. 2001).  High-speed
photometric observations from between 1997 and 2000 are described in
detail by Sokoloski \& Kenyon (2003a,b).

\begin{figure}[t]
\plotone{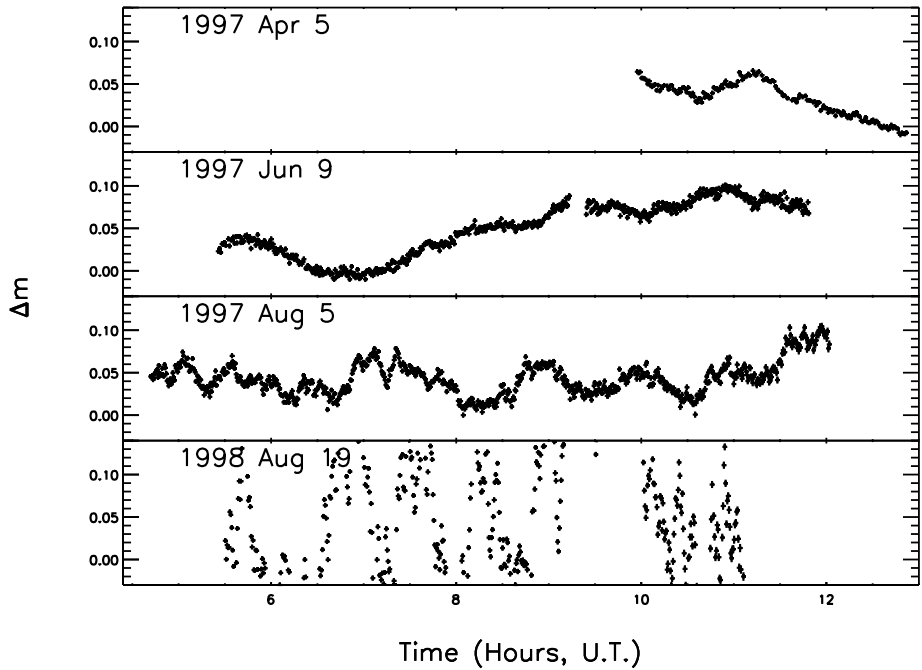}
\caption{Example CH Cyg light curves from 1997 and 1998.
The fluctuations evolve from smooth, low-amplitude variations after
the jet ejection (estimated to have occurred in mid-1996; Sokoloski \&
Kenyon 2003a), to full CV-like flickering.  The light curve from the
4th observation is off scale, since it has peak-to-peak variations of
$\sim 0.5$ mag.}
\end{figure}

The unusual, smooth light curves from 1996 and 1997 were obtained
right after material was ejected from the system and during the
development of a radio jet (Karovska, Carilli, and Mattei 1998;
Sokoloski \& Kenyon 2003a).  The bottom panel of Figure 4 shows the
times of the fast photometric observations
in the context of the rising radio flux densities.
Since both the dynamical and viscous time scales increase with
radius in the disk, and flickering is likely to be associated with
dynamical or viscous processes, 
the fastest variations generally come from the innermost portion of
the disk.  In fact, observations of stochastic and quasi-periodic
oscillations originating at different distances from the central
object in X-ray binaries, CVs, and pre-main-sequence stars support a
radial dependence of fluctuation speeds (e.g., Revnivtsev et al. 2000;
Kenyon et al. 2000; Mauche 2002).  Sokoloski \& Kenyon (2003a)
therefore interpret the absence of minute-time-scale variations in
1996 and 1997 as due to the disruption of the inner accretion disk.
CH Cyg may therefore reveal a connection between jet-producing WD
accretors and jet-producing X-ray binaries, since some X-ray binaries
have also shown evidence for a disruption of the inner accretion disk
with jet production (e.g., Feroci et al. 1999; Belloni et al. 1997).
A link between WD systems and
LMXBs with neutron-star or black-hole accretors is intriguing, since
the energetics are quite different, and the accretion flow in CH Cyg
is unlikely to be advection dominated.

\begin{figure}[t]
\plotone{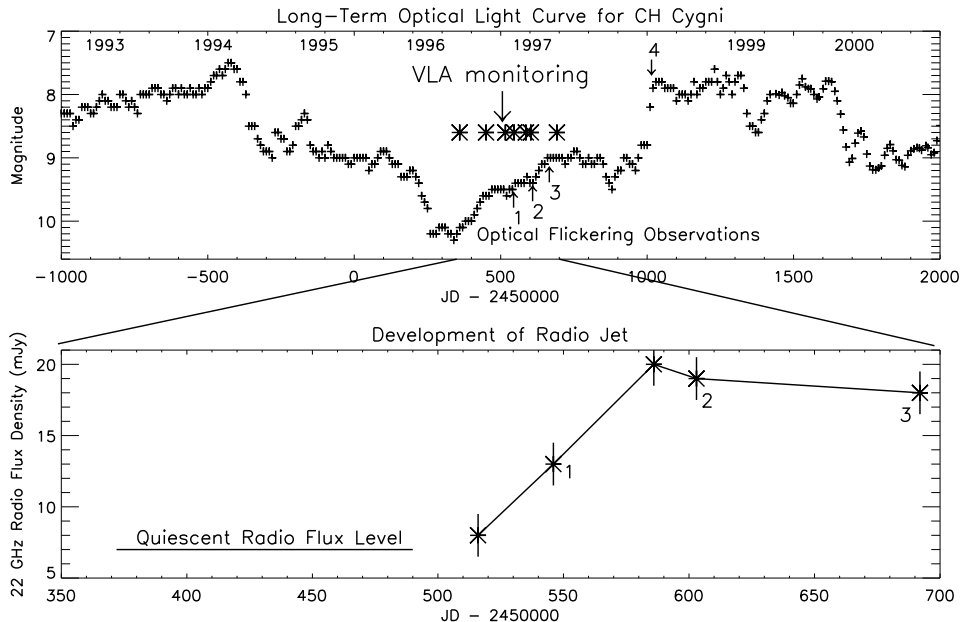}
\caption{Top panel: long-term optical light curve of CH Cyg, from the
AAVSO.  The times of the four flickering observations shown in Fig. 5
are marked with arrows, and the period of radio monitoring by Karovska
et al. (1998) is marked with stars.  Bottom panel: 22 GHz radio flux
densities, from Karovska et al. (1998).  The rise in the radio
emission was due to the production of a radio jet.}
\end{figure}

To explain the jet activity and brightness changes in CH Cyg,
Miko{\l}ajewski \& Miko{\l}ajewska (1988) proposed that the WD in CH
Cyg has a strong magnetic field, and that optical brightness changes,
flickering changes, and transient collimated jets occur when the
system moves between accretor and propeller states (see also
Miko{\l}ajewski et al. 1990a,b; Panferov \& Miko{\l}ajewski 2000;
Tomov, this volume).  In magnetic CVs, however, the optical and X-ray
emission generally oscillates at the WD spin period.  Optical
quasi-period oscillations (QPOs)
have been claimed for CH Cyg (e.g., Miko{\l}ajewski et al 1990a;
Rodgers et al. 1997), but the putative QPO frequencies are not
consistent with each other, and the significance of the claimed
detections is difficult to evaluate due to the presence of 'red noise'
in the power spectrum.  To determine the statistical significance of a
peak in a power spectrum with broad-band power, one
can assume that the noise powers are distributed
like $\chi^2$ about the best-fit power law (van der Klis 1989; Deeter
\& Boynton 1982).  Dividing the power spectrum by the best-fit
model then recovers the familiar statistical properties of white
noise, e.g., $\Pr (P_{noise} > P) = e^{-P}$ (for unbinned and
non-averaged power spectra, where $\Pr (P_{noise} > P)$ is the
probability that a statistical fluctuation will produce a peak greater
than $P$; see also Fig. 5).  Sokoloski \& Kenyon (2003b) applied this
technique to CH Cyg light curves from both the high and low states,
and did not find any significant oscillations.  They thus concluded
that alternatives to the magnetic propeller model for CH Cyg should be
considered.

\begin{figure}[t]
\plotone{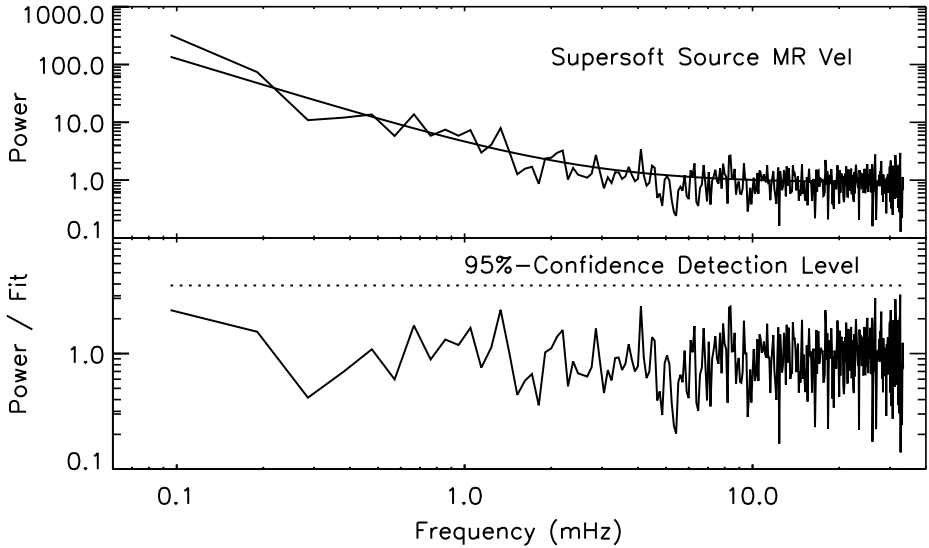}
\caption{Example of searching for an oscillation against a background
of broad-band power.  Top: Average power spectrum of SSXB MR Vel in
1995 (see \S5), and the best-fit power-law model.  Bottom: When the
power spectrum is divided by the model, standard detection statistics
can be applied, and no significant peaks are found.}
\end{figure}

As one alternative, Sokoloski \& Kenyon (2003b) suggest that the
activity in CH Cyg is driven by an unstable disk, as in dwarf novae.
The flickering amplitude in CH Cyg tends to be low when it is in an
optical low state, and high when it is in an optical high state.
Assuming that the flickering is from an accretion disk (almost all
systems with disk accretion produce stochastic brightness variations),
the correlation between flickering amplitude and optical brightness
state implies that increased mass flow through the disk is responsible
for the high states in CH Cyg.  Since jets are generally observed
after a decline in optical brightness, an unstable disk could also be
associated with (e.g., trigger or be triggered by) the outflow of
material in collimated jets.

\section{Survey of 35 SS and the Prevalence of Nuclear Shell Burning}
\label{sec:survey} 

As discussed in \S3, accretion through a disk generally produces
stochastic variations in interacting binaries.  But early flickering
studies of SS found many systems that do not flicker (Walker 1977;
Dobrzycka, Kenyon, \& Milone 1996).  Because it is not clear whether
most SS have disks, 
additional observations may help to determine whether the prevalence
of flickering is related to disk formation, or, as I argue below,
disk flickering in many SS could be hidden by emission from nuclear
burning material on the surface of the WD.

To address these questions, as well as to search for magnetic SS,
Sokoloski, Bildsten, \& Ho (1999; hereafter SBH) obtained
high-time-resolution differential CCD light curves for 35 SS, more
than doubling the number of SS observed in this way.
The light curves generally fall into three categories (excluding Z
And); Figure 6 shows examples of each type.  Twenty-five of the 35
survey objects did not vary.  In these systems, aperiodic (e.g.,
stochastic) variations were constrained to be less than just a few
mmag,
and sinusoidal oscillation amplitudes were constrained to be less than
2 - 10 mmag ($\sim$ 0.2 - 1 \% fractional variation).  Strong
stochastic variations were confirmed in five well-known flickering
systems - RS Oph, T CrB, CH Cyg, MWC 560, and Mira.  The four final
systems (EG And, BX Mon, CM Aql, and BF Cyg) varied at a low level
($\la$ tens of mmag), but need additional observations to be
considered high-confidence detections.  Thus, despite the dramatic
fast variability of a few well-studied systems like CH Cyg, and the
presence of magnetism in at least one SS (Z And), most SS do not show
either aperiodic or periodic rapid variations at a level greater than
roughly 1\%.  Furthermore, the distribution of flickering amplitudes
is bi-modal; although a few systems flicker with variations on the
order of tenths of a mag, the flickering in most systems is
constrained to less than roughly 10 mmag.

\begin{figure}[t]
\plotone{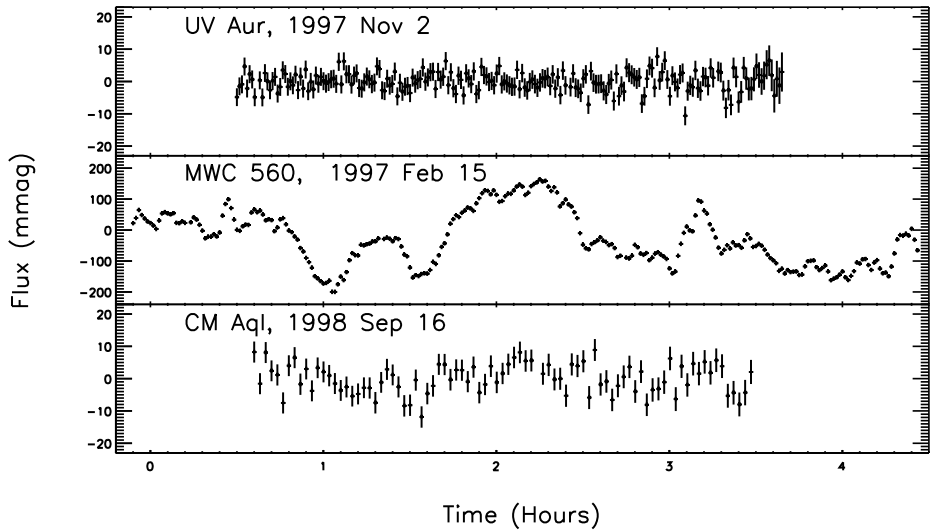}
\caption{In the survey of SBH, 25 objects had optical light curves
like UV Aur (top panel), 5 had light curves similar to MWC 560 (middle
panel), and 4 had light curves similar to CM Aql (bottom panel).}
\end{figure}

There are several possible reasons why optical flickering is absent in
most SS.  Accretion disks could be absent, or the flickering could be
hidden by emission from the red giant or luminous WD.  Because direct
evidence for an accretion disk, such as double-peaked emission lines,
has been difficult to find in SS, the hypothesis that the presence of
flickering reflects the presence of a disk is hard to test.  Optical
flickering is unlikely to be hidden by light from the red giant in
most SS because there is no systematic difference between the red
giants in the large-amplitude flickerers and the non-flickering
systems.  There is, however, a systematic difference in the WD
luminosities, $L_{hot}$.  The few low-$L_{hot}$ systems are much more
likely to flicker (as first noted by Dobrzycka et al. 1996), whereas
the high-$L_{hot}$ systems generally do not flicker.  Much of the WD
luminosity is radiated at high energies, but the nebula absorbs some
of this light and re-radiates it in the optical.  Since $E_{nuc} \gg
E_{grav}$, this nebular emission could easily hide the optical disk
flickering.  A high $L_{hot}$ indicates nuclear burning, so the
flickering amplitude seems to be related to the fundamental power
source - nuclear shell burning vs. accretion alone.  The lack of
flickering in most SS thus indicates that most SS may be powered by
nuclear shell burning on the WD.

\section{Disk Flickering from Supersoft Source MR Velorum}

As discussed in the previous section, reprocessed nuclear shell
burning can overwhelm optical disk flickering in SS.  Some SSXBs may
also to be powered by WD nuclear shell burning, so it is natural to
ask whether disk flickering is also hidden in SSXBs.  In the standard
picture, the mass-donor star in SSXBs transfers material via
Roche-lobe overflow.  An accretion disk almost certainly forms.  These
disks may be quite flared; models with high rims can reproduce eclipse
profiles in some edge-on SSXBs (Schandl, Meyer-Hofmeister, \& Meyer
1997; Meyer-Hofmeister, Schandl, \& Meyer 1997).  Dubus et al. (1999),
however, claim that an extremely flared structure is unphysical for an
illuminated disk, and that a thick inner region should shield the
outer disk.

\begin{figure}[t]
\plotone{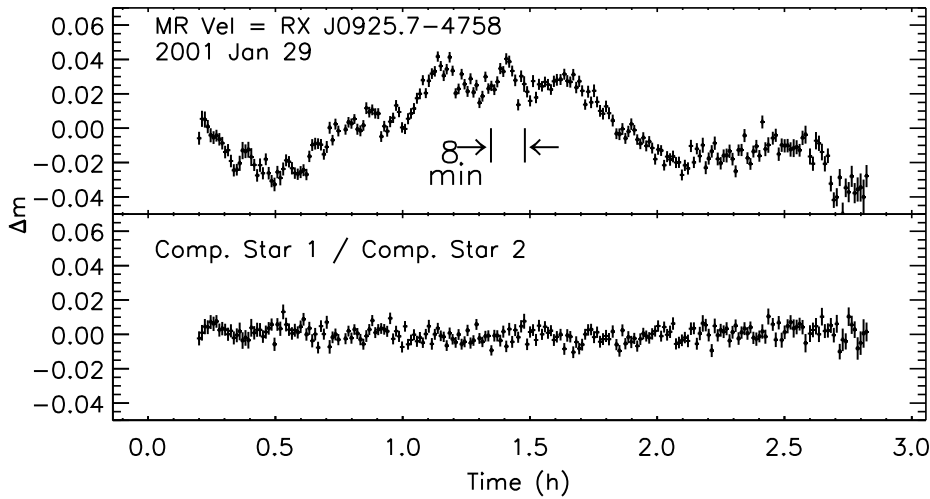}
\caption{Variations in the white-light flux of MR Vel with
respect to constant comparison stars, and ratio of comparison-star
fluxes.}
\end{figure}

In 2001 January, Sokoloski, Charles, and Clarkson (2003) observed the
Galactic transient-jet SSXB MR Vel (= RX J0925.7-4758) with the 1.9-m
telescope at SAAO.  They repeatedly found what is almost never seen in
SS -- flickering with a moderate amplitude of $\sim 60 - 80$ mmag.
Figure 7 shows one MR Vel differential white light CCD light curve.
The power spectrum is well-represented by a power-law-plus-constant
model, with a power-law slope of 1.5.  The power law drops below the
white-noise at around 4 mHz, indicating that variations with time
scales as short as 4 m were present.  Simultaneous $BVR$ light curves
are shown in Figure 8.  The overall variability amplitude was slightly
larger in $B$ ($\sigma_B = 22.1$ mmag, $\sigma_V = 16.3$ mmag, and
$\sigma_R = 17.9$ mmag), and the fastest flare was significantly
stronger in $B$ compared to $V$ and $R$.

\begin{figure}[t]
\plotone{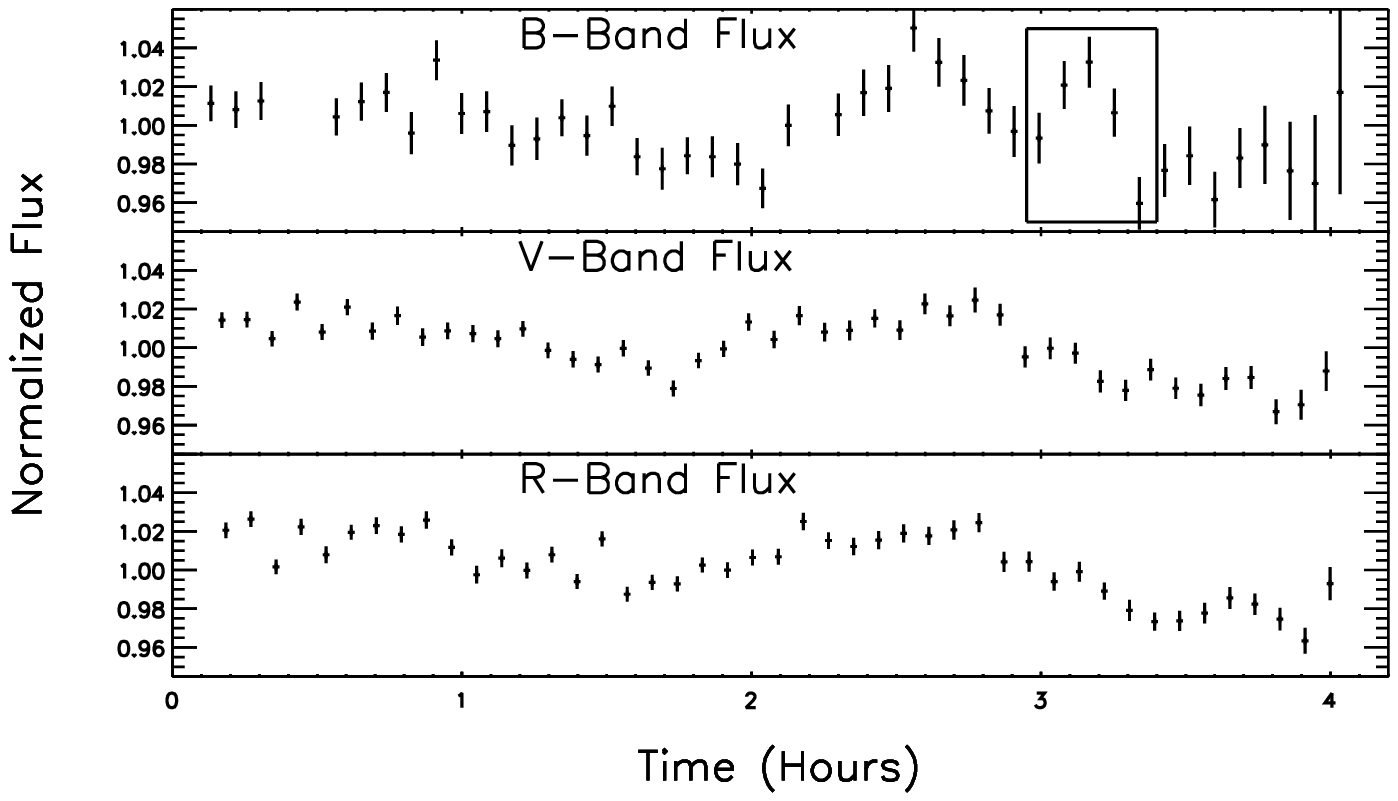}
\caption{Cyclic $BVR$ observations of SSXB MR Vel reveal that fast
flares might have bluer colors than hour-time-scale variations.}
\end{figure}

Meyer-Hofmeister et al. (1997) suggested that alterations in the size
of a reflecting disk rim (on the dynamical time at the edge of the
disk) could introduce variations into the optical emission.  If
$R_{disk} \approx 0.3 a$ (where $R_{disk}$ is the disk radius, and $a$
is the binary separation), as in CVs (Warner 1995), the disk in MR Vel
has a radius of roughly $3 \times 10^{11}\; (P_{orb}/4\, {\rm
d})^{2/3} (M_{tot} / 2\, M_{\odot})^{1/3}$ cm.  The dynamical time at
this radius (taking $M_{WD} = 1\, M_{\odot}$) is roughly 4 h.  But the
MR Vel light curves reveal variations as fast as minutes.  Therefore,
either the changing disk-rim model may not account for all the
observed variations, or the disk is smaller than a scaled-up CV disk
(i.e., $R_{disk} < 0.3 a$).

Another possible explanation for the visible flickering is that disks
in SSXBs convert less of the high-energy photons to optical photons
than are converted by the nebulae in SS.  In this case, the ratio
$L_{opt,acc}/L_{opt,nuc}$ (where $L_{opt,acc}$ is the component of the
optical luminosity from viscous dissipation in a disk and
$L_{opt,nuc}$ is the component of optical luminosity due to reflected
nuclear-burning light) could be larger in SSXBs than SS, allowing some
of the disk flickering to show through.  On the other hand, the
fractional variability in SSXBs is less than in CVs, so the there
appears to be at least some reduction in amplitude due to the nuclear
burning.  The multi-color light curves in Figure 8 support the idea
that at least some of the variable optical light in MR Vel is due to
CV-like viscous dissipation in the disk.  The bluer color of the fast
flare marked with a box in Fig. 8 compared to similar-sized slower
flares implies that the fastest flares could be due to a different
physical process (perhaps more directly related to accretion) or
originate in a different (hotter) physical region than the
hour-time-scale variations.  If they are from the inner disk, they
provide a diagnostic of the region that, as in CH Cyg, could be
associated with jet production.

\section{Fast Emission-Line Variations from RS Ophiuchi} \label{sec:rsoph}

Rapid spectral changes can in principle give us the spectral energy
distribution (SED) of the variable source, and also probe the
properties of the intervening material.  In practice, determining the
SED of the underlying flickering source is complicated by the need to
separate intrinsic source variations from Poisson variations.  Because
of the relatively lower Poisson contribution, measuring changing line
fluxes is more straight-forward.  Sokoloski et al. (2003) performed
simultaneous optical differential photometric observations and
spectroscopic observations of RS Oph on the 1- and 3-m telescopes at
UCO/Lick Observatory.  The KAST dual-armed spectrograph covered 3200 -
5300 \AA\, on the blue side, and 5300 - 6750 \AA\, on the red side.
Assuming that the red giant does not vary on a time scale of minutes,
they used normalizations from fits of a template red giant plus a
power law to the red end of each spectrum to correct for losses due to
guiding errors and seeing changes.  Figure 9 shows example plots of
the flux density (above the continuum) in several lines.

\begin{figure}[t]
\plotone{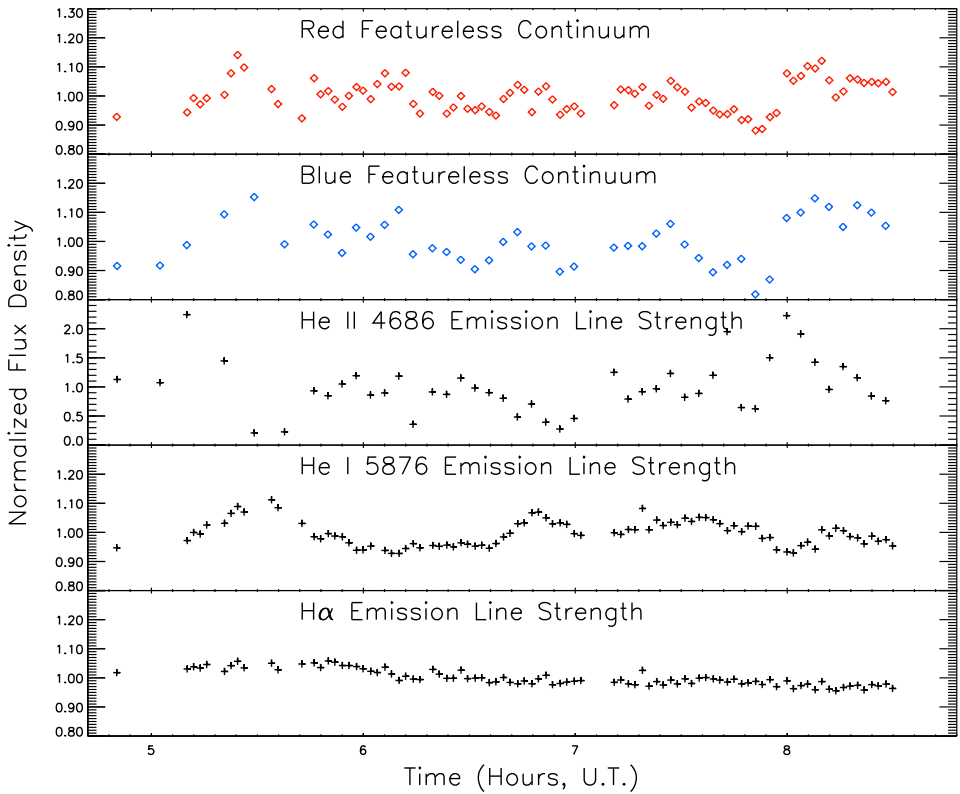}
\caption{Top two panels: flux densities from the continuum spectral
regions not contaminated by lines. They look similar to the photometric
light curve.  Middle panel: the He\,II $\lambda$4686 line strength changes in
minutes, and 
may lead the featureless-continuum variability.  Bottom two panels:
the fastest variations seen in the featureless-continuum flux density
are not present in the He\,I- and Balmer-line light curves.}
\end{figure}

The spectral flux-density variations in RS Oph prove that the red
giant was an effective internal calibration source, and they reveal
new aspects of the well-known flickering.  The light curves produced
from the featureless portions of the normalized spectra (Fig. 9, top
two panels) agree well with the simultaneous differential photometry.
The calibration technique was therefore considered reliable.  The
He\,II $\lambda$4686 line varied as fast as the blue-side observing
cadence of approximately one spectrum every 4 minutes (two red-side
spectra were taken per blue-side spectrum), and the changes were
roughly correlated with the spectral continuum and photometric
fluctuations.  Thus, there must exist a source of photons that have
energies greater than 55 eV (i.e., a source with $T \ga 50,000$ K),
such as an optically thick accretion-disk boundary layer, that varies
on a time scale of minutes.  Furthermore, the He\,II line variations
appear to lead the featureless continuum variations, suggesting that
the region which produces the He\,II emission could drive the
flickering variability.  Alternatively, some of the variable continuum
flux may be recombination radiation from the nebula (instead of, or in
addition to, optical light directly from the disk).  In this case, the
light-travel-time delay between the He\,II-line and continuum changes
would imply that the He\,II emission region is closer to the hot WD
than the continuum emission region.  The lower-ionization-state He\,I
and Balmer lines varied more slowly than the He\,II line (see Fig. 9,
bottom two panels), possibly due to the emission regions being farther
from the hot WD, longer recombination times, or a larger optical depth
in these lines.

\section{Discussion}

Rapid-variability studies suggest that the strength of aperiodic
variations in SS is related to the power source (nuclear shell burning
or accretion alone).  Systems with high $L_{hot}$ generally do not
show large-amplitude flickering, whereas SS with low $L_{hot}$ almost
always do (Walker 1977; Dobrzycka et al. 1996; SBH and references
therein).  The luminosity ratio of a typical high-$L_{hot}$ system
($\sim100 - 1000\, L_{\odot}$) to a typical low-$L_{hot}$ system
($\sim 1 - 10\, {L}_{\odot}$) is close to the ratio of the energy
released per nucleon in the nuclear burning of hydrogen-rich material
to that from accretion onto a WD.  Therefore, if nebular emission in a
symbiotic is powered by quasi-steady nuclear shell burning on the
surface of a WD, flickering or oscillations from accretion are often
hidden or reduced.  In many SSXBs, on the other hand, nuclear-burning
emission is probably reprocessed into the optical by the accretion
disk, and the ratio of reprocessed light to direct disk emission may
be low enough that some CV-like disk flickering is detectable.  Since
symbiotic recurrent novae are preferentially low-$L_{hot}$ systems,
the presence of flickering may also be related to the type of outburst
a SS experiences.

Whether a symbiotic burns material quasi-steadily or not, observations
described in \S2 and \S3 suggest that accretion-disk instabilities may
play a role in the more common, classical SS outbursts.  Furthermore,
Miko{\l}ajewska (this volume) found that SS with ellipsoidal
variations (in which the red giant is closer to filling its Roche
lobe, and a disk is more likely to form due to focusing of the red
giant wind) have more outburst activity.  So the presence of disks
could be broadly associated with outbursts in classical SS.  In CH Cyg
and also possibly in Z And (Brocksopp et al. 2003), collimated jets
are sometimes produced during or after outbursts, so disks may also be
related to the production of jets in SS.  As discussed in \S3, there
is evidence that the disk in CH Cyg was disrupted when a jet was
produced.  Similar behavior has been reported for some transient-jet
X-ray binaries, so, as suggested by Zamanov \& Marziani (2002), disks
and jets may provide a link between symbiotic and black-hole jet
sources.

Finally, periodic variations provide information about magnetism.
Given the very low oscillation amplitude in Z And, however, SBH could
not rule out strong magnetic fields in any of their survey
objects.
The detection fraction for magnetic WDs of
3\% is therefore only a crude lower limit.  
More sensitive observations are needed to determine the magnetic
fraction in SS,
and thereby test theories of the origin of magnetism in WDs and binary
stellar evolution.  Identification of additional magnetic SS is also
needed to clarify whether a strong WD field helps produce collimated
jets (as suggested by Panferov \& Miko{\l}ajewski 2000 and references
therein; Tomov, this volume) or inhibits their formation by truncating
the inner accretion disk where the jet is launched (as may be the case
in NS X-ray binaries; Fender
\& Hendry 2000).  Finally, since magnetically channeled accretion
with $\dot{M} \ga
\dot{M}_{steady,min}$ produces a large soft X-ray spin modulation, whereas
residual burning on a magnetic WD does not (King et al. 2002),
comparison between the X-ray oscillation amplitudes in magnetic SS and
magnetic SSXBs may provide information about the different (or
similar) causes of nuclear shell burning in these two classes of
systems.  

\acknowledgments

I am grateful to S. Kenyon and R. Di Stefano for helpful comments.

\end{document}